\documentclass[12pt]{article}
\usepackage{amsfonts}
\usepackage{amssymb}
\usepackage{amsmath}

\begin{document}

\title{Topological Vortices in Chiral Gauge Theory of Graphene}
\author{Xin Liu\thanks{%
Author to whom correspondence should be addressed. Electronic address:
liuxin@maths.usyd.edu.au.} \ and Ruibin Zhang \\
School of Mathematics and Statistics, University of Sydney\\
NSW 2006, Australia}
\date{ }
\maketitle

\begin{abstract}
Generation mechanism of energy gaps between conductance and valence bands is
at the centre of the study of graphene material. Recently Chamon, Jackiw, et
al. proposed a mechanism of using a Kekul\'{e} distortion background field $%
\varphi $ and its induced gauge potential $A_{i}$ to generate energy gaps.
In this paper various vortex structures inhering in this model are studied.
Regarding $\varphi $ as a generic background field rather than a fixed
Nielson-Oleson type distribution, we have found two new types of vortices on
the graphene surface --- the velocity field vortices and the monopole-motion
induced vortices --- from the inner structure of the potential $A_{i}$.
These vortex structures naturally arise from the motion of the Dirac
fermions instead of from the background distortion field.

\noindent PACS number(s): 71.10.-w, 11.15.-q, 02.40.Hw\newline
Keyword(s): Graphene; Energy Gaps; Topological Vortex Structures.
\end{abstract}

\newpage

\section{Introduction}

Graphene is a lately realized single-atomically thin $2$-dimensional sheet
of carbon. Its non-Bravais honeycomb lattice arrangement of atoms gives rise
to a new many-body problem that single-particle quantum mechanics is
relativistic, but interactions are non-relativistic instantaneous Coulomb
interactions. Ever since the breakthrough of its manufacturing technique in
2004 \cite{graphene2004} graphene has attracted enormous interest because of
the fundamental new physics exhibited and the highly unusual electronic,
mechanical and optical properties. These properties lead to numerous
potential application and make graphene a new material of growing
technological importance \cite{Geim2007NatMat}.

Graphene has a remarkable electronic property of ballistic
transport at room temperature under electrical and chemical
doping. This makes graphene a candidate to be the next dominant
semiconductor material in electronics after silicon. However, a
major obstacle for this application is that the energy spectrum of
monolayer graphene remains metallic at the Dirac points because of
the relativistic linear dispersion relation, i.e., the vanishing
gap between the conductance and valence bands in the energy
spectrum. Many methods have been developed for generating a
non-zero gap, such as spatial confinement and lateral
super-lattice potentials. In this paper we mainly consider the
generation mechanism proposed by Chamon, et al. \cite%
{ChamonPRB2000,HCM,Jackiw2007PRL,JackiwPRB2008,DHLee2009,Wilczek,Stone} and
aim to find new topological vortex structures originating from the singular
configurations of the Dirac fermion field.

The Hamiltonian for the honeycomb lattice of monolayer graphene is given by%
\begin{equation}
H_{0}=-t\sum_{\mathbf{x}\in A}\sum_{i=1}^{3}\left[ a^{\dag }\left( \mathbf{x}%
\right) b\left( \mathbf{x}+\mathbf{s}_{i}\right) +b^{\dag }\left( \mathbf{x}+%
\mathbf{s}_{i}\right) a\left( \mathbf{x}\right) \right] ,  \label{H0-rspace}
\end{equation}%
where $a$ and $b$ are respectively the fermion operators acting on the
sublattice $A$ and $B$, and $t$ the uniform hopping strength. In the
momentum space Eq.(\ref{H0-rspace}) is written as%
\begin{equation}
H_{0}=\sum_{\mathbf{k}}\left[ \tau \left( \mathbf{k}\right) a^{\dag }\left(
\mathbf{k}\right) b\left( \mathbf{k}\right) +\tau ^{\ast }\left( \mathbf{k}%
\right) b^{\dag }\left( \mathbf{k}\right) a\left( \mathbf{k}\right) \right] ,
\label{H0-pspace}
\end{equation}%
where{\ }$\left\{
\begin{array}{c}
a\left( \mathbf{k}\right) \\
b\left( \mathbf{k}\right)%
\end{array}%
\right\} =\sum_{\mathbf{x}}e^{-i\mathbf{k}\cdot \mathbf{x}}\left\{
\begin{array}{c}
a\left( \mathbf{x}\right) \\
b\left( \mathbf{x}\right)%
\end{array}%
\right\} $ and $\tau (\mathbf{k})=-t\sum_{i=1}^{3}e^{i\mathbf{k}\cdot
\mathbf{s}_{i}}$. The single-particle energy spectrum contains two Dirac
points $\mathbf{K}_{\pm }=\pm \left( \frac{4\pi }{3\sqrt{3}\ell },0\right) $
at the zone boundaries, $\ell $ being the length of lattice, satisfying $%
\tau (\mathbf{K}_{\pm })=0$. In the neighborhood of $\mathbf{K}_{\pm }$, $%
\mathbf{k}$ is written as $\mathbf{k=K_{\pm }+p}$ and $H_{0}$ is linearized
as%
\begin{eqnarray}
H_{0} &=&\sum_{\mathbf{p}}\left[ \tau _{+}\left( \mathbf{p}\right)
a_{+}^{\dag }\left( \mathbf{p}\right) b_{+}\left( \mathbf{p}\right) +\tau
_{+}^{\ast }\left( \mathbf{p}\right) b_{+}^{\dag }\left( \mathbf{p}\right)
a_{+}\left( \mathbf{p}\right) \right.  \notag \\
&&\left. +\tau _{-}\left( \mathbf{p}\right) a_{-}^{\dag }\left( \mathbf{p}%
\right) b_{-}\left( \mathbf{p}\right) +\tau _{-}^{\ast }\left( \mathbf{p}%
\right) b_{-}^{\dag }\left( \mathbf{p}\right) a_{-}\left( \mathbf{p}\right) %
\right] ,  \label{H0-linear}
\end{eqnarray}%
where $a_{\pm }\left( \mathbf{p}\right) =a\left( \mathbf{K_{\pm }+p}\right) $%
, $b_{\pm }\left( \mathbf{p}\right) =b\left( \mathbf{K_{\pm }+p}\right) $
and $\tau _{\pm }\left( \mathbf{p}\right) =\tau \left( \mathbf{K_{\pm }+p}%
\right) =\pm v_{F}\left( p_{x}\pm ip_{y}\right) $, $v_{F}$ being the Fermi
velocity, $v_{F}=1$. $H_{0}$ has the following matrix form:%
\begin{equation}
H_{0}=\Psi ^{\dag }\alpha _{i}p_{i}\Psi ,\ \ \ \ i=1,2.  \label{H0-Dirac0}
\end{equation}%
Here $\Psi \left( x\right) $ is a Dirac $4$-spinor, $\Psi \left( x\right)
=\left(
\begin{array}{cccc}
\psi _{+}^{b}\left( x\right) , & \psi _{+}^{a}\left( x\right) , & \psi
_{-}^{a}\left( x\right) , & \psi _{-}^{b}\left( x\right)%
\end{array}%
\right) ^{T}$, with $\left\{
\begin{array}{c}
a_{\pm }\left( \mathbf{p}\right) \\
b_{\pm }\left( \mathbf{p}\right)%
\end{array}%
\right\} =\int d^{2}x^{\prime }e^{i\mathbf{p}\cdot \mathbf{x}^{\prime
}}\left\{
\begin{array}{c}
\psi _{\pm }^{a}\left( \mathbf{x}^{\prime }\right) \\
\psi _{\pm }^{b}\left( \mathbf{x}^{\prime }\right)%
\end{array}%
\right\} $. The $\alpha _{i},\ i=1,2,$ is given by $\alpha _{i}=\gamma
_{4}\gamma _{i}$, accompanied by the definition of the Dirac matrices in
quantum electrodynamics: $\gamma _{k}=\left(
\begin{array}{cc}
0 & -\sigma _{k} \\
\sigma _{k} & 0%
\end{array}%
\right) $, $k=1,2,3$; $\gamma _{4}=\beta =\left(
\begin{array}{cc}
0 & I \\
I & 0%
\end{array}%
\right) $; $\gamma _{5}=i\gamma _{4}\gamma _{1}\gamma _{2}\gamma _{3}=\left(
\begin{array}{cc}
I & 0 \\
0 & -I%
\end{array}%
\right) $, $\sigma _{k}$ the Pauli matrices. The $\gamma $-matrices are
Clifford algebraic $1$-vectors, satisfying $\gamma _{a}\gamma _{b}+\gamma
_{b}\gamma _{a}=2\delta _{ab}I$, $a,b=1,2,3,4$ \cite{CliffordAlge}. From (%
\ref{H0-Dirac0}) one obtains the equation of motion for $\Psi :$%
\begin{equation}
\gamma _{a}\partial _{a}\Psi =0,\ \ \ a=1,2,4,  \label{OriDiracEqn}
\end{equation}%
which is recognized to be the massless Dirac equation in $\left( 2+1\right) $%
-dimensions. The energy dispersion of (\ref{OriDiracEqn}) is a linear
relation: $\varepsilon \left( \mathbf{p}\right) =\pm \left\vert \mathbf{p}%
\right\vert $, which has vanishing gaps between the conductance and valence
bands in the energy spectrum.

For the purpose of generating energy gaps, Chamon, Hou and Mudry \cite%
{HCM,ChamonPRB2000} introduced to the monolayer graphene a complex scalar
background field $\Delta \left( \mathbf{r}\right) $. The $\Delta \left(
\mathbf{r}\right) $ describes a Kekul\'{e} distortion of the lattice and
leads to a small variation $\delta t_{\mathbf{r},i}$ over the unit hopping
strength $t:$%
\begin{equation}
\delta t_{\mathbf{r},i}=-\frac{1}{3}\left[ \Delta \left( \mathbf{r}\right)
e^{i\mathbf{K}_{+}\cdot \mathbf{s}_{i}}e^{i\mathbf{G}\cdot \mathbf{r}%
}+\Delta ^{\ast }\left( \mathbf{r}\right) e^{-i\mathbf{K}_{+}\cdot \mathbf{s}%
_{i}}e^{-i\mathbf{G}\cdot \mathbf{r}}\right] ,\ \ \ \ \mathbf{G}=\mathbf{K}%
_{+}-\mathbf{K}_{-},  \label{del2}
\end{equation}%
$\delta t_{\mathbf{r},i}$ yielding chiral mixing between $\mathbf{K}_{+}$
and $\mathbf{K}_{-}$. With (\ref{del2}) the Hamiltonian (\ref{H0-rspace}) is
modified as%
\begin{equation}
H=-\sum_{\mathbf{r}\in A}\sum_{i=1}^{3}\left( t+\delta t_{\mathbf{r}%
,i}\right) \left[ a^{\dag }\left( \mathbf{r}\right) b\left( \mathbf{r}+%
\mathbf{s}_{i}\right) +b^{\dag }\left( \mathbf{r}+\mathbf{s}_{i}\right)
a\left( \mathbf{r}\right) \right] ,  \label{H-rspace}
\end{equation}%
and the matrix form (\ref{H0-Dirac0}) becomes%
\begin{equation}
H=\Psi ^{\dag }\mathcal{K}\Psi =\Psi ^{\dag }\left[ \mathcal{\alpha }%
_{i}p_{i}+g\beta \left( \varphi _{\mathrm{Re}}-i\gamma _{5}\varphi _{\mathrm{%
Im}}\right) \right] \Psi ,\ \ \ \ i=1,2,  \label{H-nogaupotent}
\end{equation}%
where the kernel $\mathcal{K}$ is%
\begin{equation}
\mathcal{K=}\left(
\begin{array}{cccc}
0 & -2i\partial _{z} & \Delta \left( \mathbf{r}\right) & 0 \\
-2i\partial _{\bar{z}} & 0 & 0 & \Delta \left( \mathbf{r}\right) \\
\Delta ^{\ast }\left( \mathbf{r}\right) & 0 & 0 & 2i\partial _{z} \\
0 & \Delta ^{\ast }\left( \mathbf{r}\right) & 2i\partial _{\bar{z}} & 0%
\end{array}%
\right) .  \label{H-kernel}
\end{equation}%
In (\ref{H-nogaupotent}) one denotes $\Delta \left( \mathbf{r}\right)
=g\varphi $, where $g$ is a coupling strength describing the dimension of $%
\Delta \left( \mathbf{r}\right) $, and $\varphi =\varphi _{\mathrm{Re}%
}+i\varphi _{\mathrm{Im}}$ a complex scalar, $\varphi _{\mathrm{Re}}$ and $%
\varphi _{\mathrm{Im}}$ being the real and imaginary parts respectively. $%
\varphi $ can alternatively be expressed via its modulus and phase as $%
\varphi =\left\vert \varphi \right\vert e^{i\chi }$. The energy level
equation derived from (\ref{H-nogaupotent}) reads%
\begin{equation}
\left[ \alpha _{i}p_{i}+g\beta \left( \varphi _{\mathrm{Re}}-i\gamma
_{5}\varphi _{\mathrm{Im}}\right) \right] \Psi =E\Psi .
\label{DiracEqn-mass}
\end{equation}%
It is pointed out \cite{HCM,Jackiw2007PRL} that $\chi $ can be removed by a
gauge transformation of $\varphi :$%
\begin{equation}
\chi \rightarrow \chi ^{\prime }=\chi +2\omega ,\ \ \ \ \ \ \ \Psi
\rightarrow \Psi ^{\prime }=e^{i\omega \gamma _{5}}\Psi .
\label{gaugetransf1}
\end{equation}%
Thus (\ref{H-nogaupotent}) yields $\left( \gamma _{a}\partial
_{a}+ig\left\vert \varphi \right\vert \right) \Psi ^{\prime }=0$. This gives
rise to a mass in the dispersion relation, $\varepsilon \left( \mathbf{p}%
\right) =\pm \sqrt{\mathbf{p}^{2}+\left\vert \Delta \right\vert ^{2}}$, and
hence generates a desired energy gap between the conductance and valence
bands.

For the purpose of keeping the kinetic portion of (\ref{DiracEqn-mass})
invariant against the local gauge transformation (\ref{gaugetransf1}),
Jackiw and Pi \cite{Jackiw2007PRL} extended (\ref{DiracEqn-mass}) by
introducing a $U\left( 1\right) $ gauge potential $A_{i}:$%
\begin{equation}
\left[ \alpha _{i}\left( p_{i}-\gamma _{5}A_{i}\right) +g\beta \left(
\varphi _{\mathrm{Re}}-i\gamma _{5}\varphi _{\mathrm{Im}}\right) \right]
\Psi =E\Psi ,\ \ \ \ \ \ \ \ i=1,2,  \label{pre-quant}
\end{equation}%
where $A_{i}$ transforms as $A_{i}\rightarrow A_{i}+\partial _{i}\omega $
under (\ref{gaugetransf1}). (In (\ref{pre-quant}) the chemical potential is
not considered \cite{JackiwPRB2008}.) This new model (\ref{pre-quant}) is an
equation about the Dirac fermion wavefunction $\Psi $ and the background
field $\varphi $ and its induced gauge potential $A_{i}.$ In \cite%
{Jackiw2007PRL} $\Psi $ is solved out by fixing $A_{i}$ and $\varphi $
within the Nielsen-Olesen (N-O) vortex configuration. Very recent
developments based on the above model include the study of the
superconducting vortices by P. Ghaemi, S. Ryu and D.H. Lee \cite{DHLee2009}
and P. Ghaemi and F. Wilczek \cite{Wilczek}, and the study of the
fractionally charged vortices by J.K. Pachos, M. Stone and K. Temme \cite%
{Stone}.

In this paper we will present another analysis for $\varphi $ and $A_{i}$.
Firstly, our starting point is that the background field $\varphi $ permits
more generic distributions on the graphene surface rather than a fixed N-O
type, since $\varphi $ is an externally given distortion field.
Correspondingly, we regard $A_{i}$ and $\Psi $ as two unknown quantities
that can be determined by the governing equation (\ref{pre-quant}) together
with supplementary restraints, where the restraints are naturally provided
by the subsequent study of topological vortices. Secondly, two other types
of topological vortex structures can be derived from the physical equation (%
\ref{pre-quant}) --- the velocity field vortices and the monopole-motion
induced vortices. These vortices arise from the singular configurations of
the spinor field $\Psi $ instead of from the background $\varphi $. Thirdly,
moreover, it will be shown that the potential $A_{i}$ contains a term which
describes the direct coupling of $\varphi $ and $\Psi $ fields and can be
expressed formally by an $SO\left( 4\right) $ gauge potential.

This paper is arranged as follows. In Sect.2 the inner structure of the
gauge potential $A_{i}$ in terms of the Dirac fermion spinor wavefunction $%
\Psi $ is studied. In Sect.3 two types of topological vortex structures in
the graphene system are studied. In Sect.4. the paper is summarized.

\section{Chiral Gauge Theory for Graphene}

In following it is shown that introducing the $A_{i}$ in (\ref{pre-quant})
implies formally introducing one $SO\left( 4\right) $ potential and two $%
U\left( 1\right) $ potentials.

Consider the zero-energy level of the model (\ref{pre-quant}):%
\begin{equation}
\gamma _{i}\left( \partial _{i}-i\gamma _{5}A_{i}\right) \Psi +g\left(
i\varphi _{\mathrm{Re}}+\gamma _{5}\varphi _{\mathrm{Im}}\right) \Psi =0,\ \
\ \ \ \ \ \ i=1,2.  \label{zeromodeQ}
\end{equation}%
Left-multiplying (\ref{zeromodeQ}) with $\Psi ^{\dag }\gamma _{j}\gamma _{5}$
there is%
\begin{equation}
\Psi ^{\dag }\gamma _{j}\gamma _{5}\gamma _{i}\partial _{i}\Psi +iA_{i}\Psi
^{\dag }\gamma _{j}\gamma _{i}\Psi -ig\varphi _{\mathrm{Re}}\Psi ^{\dag
}\gamma _{5}\gamma _{j}\Psi +g\varphi _{\mathrm{Im}}\Psi ^{\dag }\gamma
_{j}\Psi =0.  \label{zeromodeQ1-2}
\end{equation}%
(\ref{zeromodeQ1-2}) minus its complex conjugate gives a decomposition
expression for $A_{i}$ in terms of the spinor wavefunction $\Psi $ \cite%
{XLiuAOP08,RodriguesKeller}:%
\begin{equation}
A_{i}=A_{i}^{\left( 1\right) }+A_{i}^{\left( 2\right) }+A_{i}^{\left(
3\right) },  \label{DiracRes2-2}
\end{equation}%
where%
\begin{eqnarray}
A_{i}^{\left( 1\right) }\left( x\right) &=&g\left[ \varphi _{\mathrm{Re}}%
\frac{\Psi ^{\dag }\gamma _{5}\gamma _{i}\Psi }{\Psi ^{\dag }\Psi }+i\varphi
_{\mathrm{Im}}\frac{\Psi ^{\dag }\gamma _{i}\Psi }{\Psi ^{\dag }\Psi }\right]
,  \label{Aj(1)} \\
A_{i}^{\left( 2\right) }\left( x\right) &=&\frac{\left[ \Psi ^{\dag }\gamma
_{5}\partial _{i}\Psi -\partial _{i}\Psi ^{\dag }\gamma _{5}\Psi \right] }{%
2i\Psi ^{\dag }\Psi },  \label{Aj(2)} \\
A_{i}^{\left( 3\right) }\left( x\right) &=&\frac{\partial _{j}\left[ \Psi
^{\dag }\gamma _{5}I_{ji}\Psi \right] }{\Psi ^{\dag }\Psi },  \label{Aj(3)}
\end{eqnarray}%
$I_{ij}$ being the $SO(4)$ generator, $I_{ij}=\frac{i}{4}\left[ \gamma
_{i},\gamma _{j}\right] .$ Eq.(\ref{DiracRes2-2}) is a sufficient but not
necessary condition for the original equation (\ref{zeromodeQ}). In this
section and the next the three terms $A_{i}^{\left( 1\right) }$, $%
A_{i}^{\left( 2\right) }$ and $A_{i}^{\left( 3\right) }$ will be
investigated respectively for their topological and gauge field theoretical
essence.

$A_{i}^{\left( 1\right) }$ contains the coupling between $\Psi $ and the
background distortion field $\varphi $. The discussion is twofold:

Firstly, $A_{i}^{\left( 1\right) }$ can be formally expressed by an $%
SO\left( 4\right) $ gauge potential. Indeed, singling out the $A_{i}^{\left(
1\right) }$ part in (\ref{zeromodeQ}) leads to%
\begin{equation}
\gamma _{i}\partial _{i}\Psi -i\gamma _{i}\gamma _{5}A_{i}^{\left( 1\right)
}\Psi +ig\varphi _{\mathrm{Re}}\Psi +g\varphi _{\mathrm{Im}}\gamma _{5}\Psi
+i\gamma _{5}\gamma _{i}\left( A_{i}^{\left( 2\right) }+A_{i}^{\left(
3\right) }\right) \Psi =0,\ \ \ \ \ \ \ \ \ i=1,2,  \label{SO4rep0}
\end{equation}%
which can be compared to the following equation%
\begin{equation}
\gamma _{i}\partial _{i}\Psi -\gamma _{a}\omega _{a}\Psi +ig\varphi _{%
\mathrm{Re}}\Psi +g\varphi _{\mathrm{Im}}\gamma _{5}\Psi +i\gamma _{5}\gamma
_{i}\left( A_{i}^{\left( 2\right) }+A_{i}^{\left( 3\right) }\right) \Psi
=0,\ \ \ \ \ \ \ \ \ a=1,2,3,4.  \label{SO4rep1}
\end{equation}%
Here $\omega _{a}=\frac{1}{2}\omega _{abc}I_{bc}$ is an $SO\left( 4\right) $
potential, $\omega _{abc}$ being anti-symmetric for the $b\ c$ indices. Our
aim is to express (\ref{SO4rep0}) by (\ref{SO4rep1}), hence we write $\omega
_{abc}$ in three parts: $\omega _{abc}=\omega _{abc}^{A}+\omega
_{abc}^{S_{1}}+\omega _{abc}^{S_{2}}$, where $\omega _{abc}^{A}$ is fully
anti-symmetric for $a\;b\;c$, $\omega _{abc}^{S_{1}}$ symmetric for $a\ b$,
and $\omega _{abc}^{S_{2}}$ symmetric for $a\;c$: $\omega _{abc}^{A}=\frac{1%
}{3}(\omega _{abc}+\omega _{bca}+\omega _{cab})$, $\omega _{abc}^{S_{1}}=%
\frac{1}{3}(\omega _{abc}+\omega _{bac})$, $\omega _{abc}^{S_{2}}=\frac{1}{3}%
(\omega _{abc}+\omega _{cba})$. Defining $\bar{\omega}_{b}=2\omega _{aba},\;%
\tilde{\omega}_{a}=\epsilon _{abcd}\omega _{bcd}^{A}$, one obtains after
simple Clifford algebra: $\omega _{abc}^{A}\gamma _{a}\gamma _{b}\gamma
_{c}=i\gamma _{a}\gamma _{5}\tilde{\omega}_{a}$, $\omega
_{abc}^{S_{1}}\gamma _{a}\gamma _{b}\gamma _{c}=-\frac{2}{3}\bar{\omega}%
_{c}\gamma _{c}$, $\omega _{abc}^{S_{2}}\gamma _{a}\gamma _{b}\gamma _{c}=-%
\frac{1}{3}\bar{\omega}_{b}\gamma _{b}$. Then (\ref{SO4rep1}) becomes%
\begin{equation}
\gamma _{i}\partial _{i}\Psi -\frac{1}{4}\gamma _{a}\left( i\gamma _{5}%
\tilde{\omega}_{a}-\bar{\omega}_{a}\right) \Psi +ig\varphi _{\mathrm{Re}%
}\Psi +g\varphi _{\mathrm{Im}}\gamma _{5}\Psi +i\gamma _{5}\gamma _{i}\left(
A_{i}^{\left( 2\right) }+A_{i}^{\left( 3\right) }\right) \Psi =0.
\label{SO4rep2}
\end{equation}%
Thus comparing (\ref{SO4rep2}) with (\ref{SO4rep0}) one arrives at a
conclusion that if the evaluation%
\begin{equation}
\bar{\omega}_{a}=0;\ \ \ \tilde{\omega}_{i}=4A_{i}^{\left( 1\right) },\
i=1,2;\ \ \ \tilde{\omega}_{3}=\tilde{\omega}_{4}=0  \label{Wevaluation}
\end{equation}%
is taken, the $2$-component potential $A_{i}^{\left( 1\right) },\ i=1,2,$
can be formally expressed by the $4$-component $SO\left( 4\right) $
potential $\omega _{a},\ a=1,2,3,4,$\ via the expression (\ref{SO4rep1}).
This is the gauge field theoretical essence of the potential $A_{i}^{\left(
1\right) }$. It is addressed in addition that since $A_{i}^{\left( 1\right)
} $ is real and $\gamma _{i}$ anti-Hermitian, the $SO(4)$ potential $\omega
_{a}$ is required to be Hermitian, which is different from ordinary
anti-Hermitian gauge potentials in physics. For the discussion of Hermitian
gauge potentials see \cite{ImPotent1}.

Secondly, $A_{i}$ is the induced gauge potential of the background
distortion $\varphi $, and $A_{i}^{\left( 1\right) }$ the only term of $%
A_{i} $ which explicitly contains $\varphi $. Our second observation for $%
A_{i}^{\left( 1\right) }$ is that $\varphi $ enters the potential $A_{i}$ by
coupling to the Clifford algebraic odd vectors \cite{CliffordAlge}. Go back
to (\ref{zeromodeQ}) and write the $A_{i}^{\left( 1\right) }$ part explicitly%
\begin{eqnarray}
\gamma _{i}\partial _{i}\Psi &+&ig\varphi _{\mathrm{Re}}\frac{\Psi ^{\dag
}\gamma _{5}\gamma _{i}\Psi }{\Psi ^{\dag }\Psi }\gamma _{5}\gamma _{i}\Psi
+g\varphi _{\mathrm{Im}}\frac{\Psi ^{\dag }\gamma _{i}\Psi }{\Psi ^{\dag
}\Psi }\gamma _{i}\gamma _{5}\Psi  \notag \\
&+&i\gamma _{5}\gamma _{i}\left( A_{i}^{\left( 2\right) }+A_{i}^{\left(
3\right) }\right) \Psi +ig\varphi _{\mathrm{Re}}\Psi +g\varphi _{\mathrm{Im}%
}\gamma _{5}\Psi =0,\ \ \ \ \ \ \ \ \ \ \ i=1,2.  \label{Aj1-a}
\end{eqnarray}%
Keep in mind that $\gamma _{i}$ is a Clifford algebraic $1$-vector and $%
\gamma _{5}\gamma _{i}$ a $3$-vector (or pseudo $1$-vector). Then
introducing the projection operator $\hat{P}=\frac{\Psi \Psi ^{\dag }}{\Psi
^{\dag }\Psi }$, Eq.(\ref{Aj1-a}) reads%
\begin{equation}
\gamma _{i}\partial _{i}\Psi +g\left( i\varphi _{\mathrm{Re}}\hat{P}_{\gamma
_{5}\gamma _{i}}+\varphi _{\mathrm{Im}}\hat{P}_{\gamma _{i}}\gamma
_{5}\right) \Psi +i\gamma _{5}\gamma _{i}\left( A_{i}^{\left( 2\right)
}+A_{i}^{\left( 3\right) }\right) \Psi +ig\varphi _{\mathrm{Re}}\Psi
+g\varphi _{\mathrm{Im}}\gamma _{5}\Psi =0.  \label{oddCliffspace}
\end{equation}%
Here $\hat{P}_{\gamma _{5}\gamma _{i}}=Tr\left( \hat{P}\gamma _{5}\gamma
_{i}\right) \gamma _{5}\gamma _{i}$ and $\hat{P}_{\gamma _{i}}=Tr\left( \hat{%
P}\gamma _{i}\right) \gamma _{i}$ are the Clifford vectorial components of $%
\hat{P}$ respectively along the $3$-vector $\gamma _{5}\gamma _{i}$ and the $%
1$-vector $\gamma _{i}$. Hence it is concluded from (\ref{oddCliffspace})
that $\varphi $ enters the potential $A_{i}$ by coupling to the Clifford odd
vectors.

Moreover, it can be checked that $A_{i}^{\left( 1\right) }$ and $%
A_{i}^{\left( 3\right) }$ are invariant under the gauge transformations (\ref%
{gaugetransf1}) of $\varphi $, whereas $A_{i}^{\left( 2\right) }$ undergoes
a $U\left( 1\right) $-type gauge potential transformation: $A_{i}^{\left(
1\right) }\rightarrow A_{i}^{\left( 1\right) \prime }=A_{i}^{\left( 1\right)
}$, $A_{i}^{\left( 3\right) }\rightarrow A_{i}^{\left( 3\right) \prime
}=A_{i}^{\left( 3\right) };\ \ \ \ A_{i}^{\left( 2\right) }\rightarrow
A_{i}^{\left( 2\right) \prime }=A_{i}^{\left( 2\right) }+\partial _{i}\omega
.$

\section{Topological Vortex Structures}

In this section the $A_{i}^{\left( 2\right) }$ and $A_{i}^{\left( 3\right) }$
terms of (\ref{DiracRes2-2}) will be investigated. It will be shown that
they lead to different types of topological vortices on the graphene surface.

\subsubsection*{Vortices Arising from $A_{i}^{\left( 2\right) }$}

With respect to the chiral $\gamma _{5}$ matrix one considers the self-dual
and anti-self-dual part of $\Psi $ separately: $\Psi =\Psi _{+}+\Psi _{-}$,
with $\Psi _{\pm }=\frac{1}{2}\left( 1\pm \gamma _{5}\right) \Psi $
satisfying $\gamma _{5}\Psi _{\pm }=\pm \Psi _{\pm }$. The explicit
representation of $\Psi _{\pm }$ is: $\Psi _{+}=\left(
\begin{array}{cc}
\Phi _{+}^{T} & 0%
\end{array}%
\right) ^{T}$, $\Psi _{-}=\left(
\begin{array}{cc}
0 & \Phi _{-}^{T}%
\end{array}%
\right) ^{T}$, where $\Phi _{+}\left( x\right) $ and $\Phi _{-}\left(
x\right) $ are two Pauli $2$-spinors and have their respective $SU\left(
2\right) $ sub-group spaces (hereafter marked as $SU\left( 2\right) _{\pm }$
subspaces). Thus (\ref{Aj(2)}) reads%
\begin{equation}
A_{i}^{\left( 2\right) }=\cos ^{2}\theta \frac{\Phi _{+}^{\dag }\partial
_{i}\Phi _{+}-\partial _{i}\Phi _{+}^{\dag }\Phi _{+}}{2i\Phi _{+}^{\dag
}\Phi _{+}}-\sin ^{2}\theta \frac{\Phi _{-}^{\dag }\partial _{i}\Phi
_{-}-\partial _{i}\Phi _{-}^{\dag }\Phi _{-}}{2i\Phi _{-}^{\dag }\Phi _{-}},
\end{equation}%
where $\cos ^{2}\theta =\frac{\Psi _{+}^{\dag }\Psi _{+}}{\Psi ^{\dag }\Psi }
$ and $\sin ^{2}\theta =\frac{\Psi _{-}^{\dag }\Psi _{-}}{\Psi ^{\dag }\Psi }
$, $\theta \left( x\right) $ called the duality rotation \cite{XLiuAOP08}.

As mentioned in the introductory section, in the present paper the gauge
potential $A_{i}$ and the spinor wave function $\Psi $ are regarded as two
unknowns determined by the governing equation (\ref{pre-quant}), with the
background field $\varphi $ being a generic externally given distortion for
the graphene system. (\ref{pre-quant}) provides $8$ restraints but $A_{i}$
and $\Psi $\ have totally $10$ unknown components. Hence for determining the
unknowns one needs two extra supplementary restraints. In this regard one
employs the following normalization conditions of $\Psi $ to play the role
of the required supplementary restraints:%
\begin{equation}
\sin ^{2}\theta =\cos ^{2}\theta =\frac{1}{2},\ \ \ \ \ \ \Psi ^{\dag }\Psi
=1.  \label{supplrestr}
\end{equation}%
Eq.(\ref{supplrestr}) means that there is no priority between the self- and
anti-self-dual subspaces, hence (\ref{supplrestr}) is a natural choice. Thus%
\begin{equation}
A_{i}^{\left( 2\right) }=\frac{1}{2}\sum_{\pm }\pm A_{i\pm }^{\left(
2\right) },\ \ \ \ \ \ A_{i\pm }^{\left( 2\right) }=\frac{1}{i}\left[ \Phi
_{\pm }^{\dag }\partial _{i}\Phi _{\pm }-\partial _{i}\Phi _{\pm }^{\dag
}\Phi _{\pm }\right] .  \label{A(1)velvort}
\end{equation}%
The bilinear form $A_{i+}^{\left( 2\right) }\left( x\right) $ [resp. $%
A_{i-}^{\left( 2\right) }\left( x\right) $] is a $U\left( 1\right) $ gauge
potential in the $SU\left( 2\right) _{+}$ [resp. $SU\left( 2\right) _{-}$]
subspace, and has the physical meaning of the velocity field of spinning
quantum fluid, up to a mass constant.

Topologically singular property of the configuration of $\Phi _{\pm }$ is
studied by considering the first Chern class, a topological characteristic
class on the $2$-dimensional graphene surface \cite{Nash}. The first Chern
class constructed from $A_{i\pm }^{\left( 2\right) }$ is $C_{1\pm }^{\left(
2\right) }=\frac{1}{2\pi }\frac{1}{2}f_{ij\pm }^{\left( 2\right)
}dx^{i}\wedge dx^{j}$, where $f_{ij\pm }^{\left( 2\right) }$ is the gauge
field strength defined as $f_{ij\pm }^{\left( 2\right) }=\partial
_{i}A_{j\pm }^{\left( 2\right) }-\partial _{j}A_{i\pm }^{\left( 2\right) }$.
The physical meaning of $C_{1\pm }^{\left( 2\right) }$ is the vorticity of
the velocity field distributed on the graphene surface. It is known in
hydrodynamics that non-vanishing vorticity possesses vortices; this fact can
be geometrically shown here for $C_{1\pm }^{\left( 2\right) }$. Indeed, a
spinning quantum fluid such as the Helium-3 superfluid has the so-called
Mermin-Ho relation \cite{velvor}%
\begin{equation}
\partial _{i}A_{j\pm }^{\left( 2\right) }-\partial _{j}A_{i\pm }^{\left(
2\right) }=\frac{1}{2}\epsilon _{abc}n_{\pm }^{a}\partial _{i}n_{\pm
}^{b}\partial _{j}n_{\pm }^{c}d^{2}x,  \label{A(1)vort}
\end{equation}%
where $n_{\pm }^{a}$, $a=1,2,3$, denote the spin unit vectors respectively
in the\ $SU\left( 2\right) _{\pm }$\ subspaces, $n_{\pm }^{a}\left( x\right)
=\frac{\Phi _{\pm }^{\dag }\sigma _{a}\Phi _{\pm }}{\Phi _{\pm }^{\dag }\Phi
_{\pm }}$. From the geometric point of view, the RHS of (\ref{A(1)vort}) is
the pullback of a surface element on the hypersphere $S_{\pm }^{2}$, where $%
S_{\pm }^{2}$ is formed by $n_{\pm }^{a}$ in the group space of $SU\left(
2\right) _{\pm }$. The $S_{\pm }^{2}$ tangentially intersects the graphene
surface at the studied point $\mathbf{x}=\left( x^{1},x^{2}\right) $, hence $%
n_{\pm }^{a}$ is locally perpendicular to the graphene surface at $\mathbf{x}
$. Meanwhile, it is known that the $S_{\pm }^{2}$ surface element can be
geometrically given by a $U\left( 1\right) $-type gauge potential $w_{\pm }$
as%
\begin{equation}
\epsilon _{abc}n_{\pm }^{a}dn_{\pm }^{b}dn_{\pm }^{c}=dw_{\pm }.
\label{wu-yangpot}
\end{equation}%
Here $w_{\pm }$ is defined as $w_{\pm }=\vec{e}_{1\pm }\cdot d\vec{e}_{2\pm
} $, where $\vec{e}_{1+}$\ and $\vec{e}_{2+}$\ [resp. $\vec{e}_{1-}$\ and $%
\vec{e}_{2-}$] are a pair of $2$-dimensional unit vectors normal to $%
n_{+}^{a}$ [resp. $n_{-}^{a}$] on $S_{+}^{2}$ [resp. $S_{-}^{2}$]. Namely, $%
\left( \vec{e}_{1+},\;\vec{e}_{2+},\;\vec{n}_{+}\right) $\ [resp. $\left(
\vec{e}_{1-},\;\vec{e}_{2-},\;\vec{n}_{-}\right) $] forms an orthogonal
frame: $\vec{e}_{\pm 1}\cdot \vec{e}_{\pm 2}=\vec{e}_{\pm 1}\cdot \vec{n}%
_{\pm }=\vec{e}_{\pm 2}\cdot \vec{n}_{\pm }=0$, $\vec{e}_{\pm 1}\cdot \vec{e}%
_{\pm 1}=\vec{e}_{\pm 2}\cdot \vec{e}_{\pm 2}=\vec{n}_{\pm }\cdot \vec{n}%
_{\pm }=1$. The $w_{\pm }$\ has the physical meaning of the so-called
Wu-Yang potential \cite{WuYang}. Comparing (\ref{A(1)vort}) and (\ref%
{wu-yangpot}) one knows that locally%
\begin{equation}
A_{i\pm }^{\left( 2\right) }=w_{i\pm }=\vec{e}_{1\pm }\cdot \partial _{i}%
\vec{e}_{2\pm }  \label{wu-yangpot2}
\end{equation}%
up to a removable phase angle. Hence in following one can use (\ref%
{wu-yangpot2}) to investigate the topological structures arising from $%
A_{i\pm }^{\left( 2\right) }$.

For convenience we use one unique $2$-component vector to replace the pair $%
\left( \vec{e}_{1\pm },\vec{e}_{2\pm }\right) $. This vector, denoted as $%
\vec{\xi}_{\pm }=(\xi _{\pm }^{1},\xi _{\pm }^{2})$, is required to reside
in the plane spanned by $\vec{e}_{1\pm }$ and $\vec{e}_{2\pm }$ and satisfy $%
e_{1\pm }^{A}=\frac{\xi _{\pm }^{A}}{\left\Vert \xi _{\pm }\right\Vert }%
,\;e_{2\pm }^{A}=\epsilon ^{AB}\frac{\xi _{\pm }^{B}}{\left\Vert \xi _{\pm
}\right\Vert },$ with $\left\Vert \xi _{\pm }\right\Vert ^{2}=\xi _{\pm
}^{A}\xi _{\pm }^{A}$ and $A,B=1,2$. The zero points of $\vec{\xi}_{\pm }$
are the singular points of $\vec{e}_{1\pm }$ and $\vec{e}_{2\pm }$. Then in
terms of $\vec{\xi}_{\pm }$ one rewrites $A_{i\pm }^{\left( 2\right)
}=\epsilon _{AB}\frac{\xi _{\pm }^{A}}{\left\Vert \xi _{\pm }\right\Vert }%
\partial _{i}\frac{\xi _{\pm }^{B}}{\left\Vert \xi _{\pm }\right\Vert },$
and then obtains $C_{1\pm }^{\left( 2\right) }=\frac{1}{2\pi }\epsilon
_{ij}\epsilon _{AB}\partial _{i}\frac{\xi _{\pm }^{A}}{\left\Vert \xi _{\pm
}\right\Vert }\partial _{j}\frac{\xi _{\pm }^{B}}{\left\Vert \xi _{\pm
}\right\Vert }d^{2}x$. According to \cite{XLiuAOP} it can be proved that%
\begin{equation}
C_{1\pm }^{\left( 2\right) }=\delta ^{2}\left( \vec{\xi}_{\pm }\right) D(%
\frac{\xi _{\pm }}{x})d^{2}x,  \label{2dimtopcurr}
\end{equation}%
where $D(\xi _{\pm }/x)=\frac{1}{2}\epsilon ^{ij}\epsilon _{AB}\partial
_{i}\xi _{\pm }^{A}\partial _{j}\xi _{\pm }^{B}$ is the Jacobian
determinant. Eq.(\ref{2dimtopcurr}) shows that non-vanishing $C_{1\pm
}^{\left( 2\right) }$ occurs only at the zero-points of $\vec{\xi}_{\pm }$.
Hence in order to find the expected topological vortices on the graphene
surface one should study the zero point equation of $\vec{\xi}_{\pm }$: $\xi
^{A}(x)=0$, $A=1,2$. The implicit function theory \cite{ImplicitFunc}
declares that under the regular condition $D\left( \xi _{\pm }/x\right) \neq
0$ the general solutions of the zero point equations are a finite number of $%
2$-dimensional isolated points:%
\begin{equation}
x_{\pm }^{i}=x_{k_{\pm }}^{i},\ \ \ \ \ \ k_{\pm }=1,2,...,L_{\pm },
\end{equation}%
where $L_{\pm }$ denotes the number of the isolated points. Thus these
points are the predicted topological singularities in the configuration of
the $\Phi _{\pm }$ field on the graphene surface. In the scenario of
hydrodynamics they are called the velocity field vortices, the topological
vortices arising from non-vanishing vorticity.

The $\delta ^{2}\left( \vec{\xi}_{\pm }\right) $ in
(\ref{2dimtopcurr}) can be expanded onto these vortex points as\\
$\delta ^{2}\left( \vec{\xi}_{\pm
}\right) =\sum_{k_{\pm }=1}^{L_{\pm }}W_{k_{\pm }}\delta ^{2}\left( \vec{x}-%
\vec{x}_{k_{\pm }}\right) ,$ where $W_{k_{\pm }}$ is the winding number of
the $k_{\pm }$-th zero-point, playing the role of the topological charge of
that zero-point. Therefore the first Chern number given by the Chern class $%
C_{1\pm }^{\left( 2\right) }$ is%
\begin{equation}
c_{1\pm }^{\left( 2\right) }=\int C_{1\pm }^{\left( 2\right) }=\sum_{k_{\pm
}=1}^{L_{\pm }}W_{k_{\pm }}.
\end{equation}%
$c_{1\pm }^{\left( 2\right) }$ has the physical meaning of quantized
vorticity.

\subsubsection*{Vortices Arising from $A_{i}^{\left( 3\right) }$}

Another kind of vortex structures originate from the third term of Eq.(\ref%
{DiracRes2-2}), $A_{i}^{\left( 3\right) }=\partial _{j}\left[ \Psi ^{\dag
}\gamma _{5}I_{ji}\Psi \right] $. Defining the dual tensor of $I_{ij}$ as $%
^{\ast }I_{ij}=\gamma _{5}I_{ij}=\frac{1}{2}\epsilon _{ijkl}I_{kl}\ \ \left(
i,j=1,2;\ k,l=3,4\right) $, one has
\begin{equation}
A_{i}^{\left( 3\right) }=\partial _{j}\left[ \Psi ^{\dag }\phantom{,}^{\ast
}I_{ji}\Psi \right] .  \label{A(2)dual1}
\end{equation}%
The RHS of (\ref{A(2)dual1}) can be expressed by a Maxwell-type $U\left(
1\right) $ field tensor. Indeed, according to \cite{RodriguesKeller} a
Maxwell-type $U\left( 1\right) $ electromagnetic field strength $F_{\mu \nu
} $ has a Dirac spinor representation $F_{\mu \nu }=\Psi ^{\dag }I_{\mu \nu
}\Psi ,\ \mu ,\nu =1,2,3,4,$ provided that $\Psi $ is non-singular.
Therefore, for (\ref{A(2)dual1}) one can extend the studied base space from
the $2$-dimensional to the $4$-dimensional, and consider only the weak
solutions for $\Psi $. Here the weak solutions refer to the case that the $%
\Psi $ field has only a countable number of isolated singular points and is
well-defined almost everywhere except at the singular points. Then the RHS
of (\ref{A(2)dual1}) can be expressed by a dual electromagnetic field
strength as $A_{i}^{\left( 3\right) }=\partial _{j}\phantom{,}^{\ast }F_{ji}$%
, where $^{\ast }F_{ij}=\frac{1}{2}\epsilon _{ijkl}F_{kl}$ is the dual
tensor of $F_{ij}$.

The second Maxwell equation reads $\partial _{j}\phantom{,}^{\ast
}F_{ji}=-4\pi \phantom{,}^{\ast }J_{i}$, where $^{\ast }J_{i}$ is the
current of monopoles. When $^{\ast }J_{i}=0,$ namely there is no existence
of monopole-type excitations in the system, there is $\partial _{j}%
\phantom{,}^{\ast }F_{ji}=0,$ which corresponds to the Bianchi identity. In
this case, $A_{i}^{\left( 3\right) }=0$; When $^{\ast }J_{i}\neq 0$, $%
A_{i}^{\left( 3\right) }=-4\pi \phantom{,}^{\ast }J_{i}$. One needs to
consider the vortex structures arising from the motion of monopoles on the $%
2 $-dimensional graphene surface.

There are various models for magnetic monopoles in literature \cite%
{Nash,monopoleliterature}, depending on different choices for the symmetry
of the monopole field or the gauge potential. In following we simply
consider the complex scalar field model for the monopoles, and let the
monopole current and density take the following form%
\begin{equation}
^{\ast }J_{i}=\frac{1}{2i\psi ^{\ast }\psi }\left( \psi ^{\ast }\partial
_{i}\psi -\partial _{i}\psi ^{\ast }\psi \right) ,\ \ i=1,2;\ \ \ \ \ \ \rho
=\psi ^{\ast }\psi ,  \label{monoU1}
\end{equation}%
where $\psi $ is a complex scalar, $\psi =\phi ^{1}+i\phi ^{2}$. (\ref%
{monoU1}) leads to $^{\ast }J_{i}=\epsilon ^{\alpha \beta }\left( \frac{\phi
^{\alpha }}{\left\vert \phi \right\vert }\right) \partial _{i}\left( \frac{%
\phi ^{\beta }}{\left\vert \phi \right\vert }\right) $ with $\left\vert \phi
\right\vert ^{2}=\phi ^{\alpha }\phi ^{\alpha },\ \alpha ,\beta =1,2$. The
first Chern class induced by $A_{i}^{\left( 3\right) }$ is $C_{1}^{\left(
3\right) }=\frac{1}{2}\epsilon ^{\alpha \beta }\epsilon ^{ij}\partial
_{i}\left( \frac{\phi ^{\alpha }}{\left\vert \phi \right\vert }\right)
\partial _{j}\left( \frac{\phi ^{\beta }}{\left\vert \phi \right\vert }%
\right) .$ Since only the case of weak solutions is taken for $\Psi $, $%
C_{1}^{\left( 3\right) }$ can be expressed in the $\delta $-function form as
$C_{1}^{\left( 3\right) }=\delta ^{2}\left( \vec{\phi}\right) D\left( \frac{%
\phi }{x}\right) $, similarly to the last subsection. Apparently
non-vanishing $C_{1}^{\left( 3\right) }$ occurs only at the zero-points of $%
\vec{\phi}$, so the zero point equation $\phi ^{\alpha }(x)=0$, $\alpha =1,2$%
, should be studied. Similarly, under the regular condition $D\left( \phi
/x\right) \neq 0$ the general solutions of the zero-point equations are a
finite number of $2$-dimensional isolated points:%
\begin{equation}
x^{i}=x_{\ell }^{i},\ \ \ \ \ \ell =1,2,...,N,
\end{equation}%
where $N$ denotes the number of the isolated points. These points are
topological singularities in the configuration of the $\phi $ field on the
graphene surface; they are vortex structures due to the motion of monopoles.
Furthermore, the $\delta ^{2}\left( \vec{\phi}\right) $ can be expanded onto
the vortex points as $\delta ^{2}\left( \vec{\phi}\right) =\sum_{\ell
=1}^{N}W_{\ell }\delta ^{2}\left( \vec{x}-\vec{x}_{\ell }\right) $, where $%
W_{\ell }$ is the winding number of the $\ell $-th zero-point, playing the
role of the topological charge of that zero-point. Correspondingly the first
Chern number given by the Chern class $C_{1}^{\left( 3\right) }$ is%
\begin{equation}
c_{1}^{\left( 3\right) }=\int C_{1}^{\left( 3\right) }=\sum_{\ell
=1}^{N}W_{\ell }.
\end{equation}

\section{Conclusion}

The massive Dirac equation (\ref{zeromodeQ}) proposed in Ref.\cite%
{Jackiw2007PRL} is an important mechanism for generating energy gaps between
conductance and valence bands of graphene material. We start from this
equation and proposed that the distortion field $\varphi $ is permitted to
take a generic distribution rather than a fixed Nielson-Oleson (N-O)
configuration on the graphene surface. The $\varphi $-induced gauge
potential $A_{i}$ and the Dirac fermion wavefunction $\Psi $ are determined
by (\ref{zeromodeQ}) and the normalization condition (\ref{supplrestr}). In
this paper we focus on the inner spinor structure of $A_{i}$ and study its
three terms $A_{i}^{\left( 1\right) },\ A_{i}^{\left( 2\right) }$ and $%
A_{i}^{\left( 3\right) }$. Emphasis is on two new types of topological
vortices arising from $A_{i}^{\left( 2\right) }$ and $A_{i}^{\left( 3\right)
}$ --- the velocity field vortices and the monopole-motion induced vortices.
These vortices are different from the N-O vortices, as they both originate
from the singularities of the wavefunction $\Psi $ instead of from the
background field $\varphi $. Moreover it is shown that the $A_{i}^{\left(
1\right) }$ term can be formally expressed via an $SO\left( 4\right) $ gauge
potential $\omega _{a}$.

\section{Acknowledgement}

The author X.L. is indebted to Profs. J. Keller and W.A. Rodrigues Jr. for
the discussion on the equivalence between the Maxwell and Dirac equations.
X.L. was financially supported by the USYD Postdoctoral Fellowship of the
University of Sydney. R.Z. was financially supported by the Australian
Research Council.

\end{document}